\documentclass[aps,prl,twocolumn,superscriptaddress]{revtex4}
\usepackage{amsmath}
\usepackage{graphicx}
\begin{document}
\draft
\preprint{}

\title{\bf Temperature F\/luctuations of the Cosmic Microwave
Background Radiation: \\
A Case of Nonextensivity?}

\author{Armando Bernui}
\email[]{bernui@das.inpe.br}
\altaffiliation{on leave from: Facultad de Ciencias,
Universidad Nacional de Ingenier\'{\i}a, Lima, Peru}
\affiliation{Instituto Nacional de Pesquisas Espaciais,
             Divis\~{a}o de Astrof\'{\i}sica  \\
Av. dos Astronautas 1758, 12227-010 -- S\~ao Jos\'e dos Campos,
SP, Brazil}

\author{Constantino Tsallis}
\email[]{tsallis@cbpf.br}
\affiliation{Centro Brasileiro de Pesquisas F\'{\i}sicas \\
Rua X. Sigaud 150, 22290-180 -- Rio de Janeiro, RJ, Brazil}
\affiliation{Santa Fe Institute, 1399 Hyde Park Road, Santa Fe,
New Mexico 87501, USA}

\author{Thyrso Villela}
\email[]{thyrso@das.inpe.br}
\affiliation{Instituto Nacional de Pesquisas Espaciais,
             Divis\~{a}o de Astrof\'{\i}sica  \\
Av. dos Astronautas 1758, 12227-010 -- S\~ao Jos\'e dos Campos,
SP, Brazil}

\begin{abstract}
\noindent 
Temperature maps of the Cosmic Microwave Background (CMB) radiation, as 
those obtained by the Wilkinson Microwave Anisotropy Probe (WMAP), provide 
one of the most precise data sets to test fundamental hypotheses of modern 
cosmology. 
One of these issues is related to the statistical properties of the CMB 
temperature f\/luctuations, which would have been produced by Gaussian 
random density f\/luctuations when matter and radiation were in thermal 
equilibrium in the early Universe. 
We analysed here the WMAP data and found that the distribution of the CMB 
temperature f\/luctuations $P^{\text{CMB}}(\Delta T)$ can be quite well 
fitted by the anomalous temperature distribution emerging within nonextensive 
statistical mechanics. 
This theory is based on the nonextensive entropy 
$S_q \equiv k \{ 1-\int dx \, [P_q(x)]^q \} /(q-1)$, with the Boltzmann-Gibbs 
expression as the limit case $q \to 1$. 
For the frequencies investigated ($\nu=$ 40.7, 60.8, and 93.5 GHz), we found 
that $P^{\text{CMB}}(\Delta T)$ is well described by 
$P_q(\Delta T) \propto 1/[1+(q-1) B(\nu) \Delta T\:\!^2]^{1/(q-1)}$, with 
$q = 1.055 \pm 0.002$,  
which exclude, at the 99\% confidence level, exact Gaussian temperature 
distributions $P^{\text{Gauss}}(\Delta T) \propto e^{- B(\nu) \Delta T\:\!^2}$, 
corresponding to the $q \to 1$ limit, to properly represent the CMB 
temperature f\/luctuations measured by WMAP.
\end{abstract}

\date{\today}

\pacs{98.80.-k;98.70.Vc;05.90.+m}

\maketitle

\noindent
\small{{\textbf Keywords:} Cosmic Background Radiation: temperature 
f\/luctuations; Nonextensive statistical mechanics; Nongaussian distributions.}

Cosmic Microwave Background (CMB) radiation is formed by the leftover photons 
from the early Universe, when matter and radiation were coupled in thermal 
equilibrium. 
With the expansion of the Universe, these photons decoupled and spread out 
freely throughout space, basically conserving their primordial features. 
In the early 90's, the Far Infrared Absolute Spectrophotometer, on board 
the Cosmic Background Explorer (COBE) satellite, proved that this radiation 
was in thermal equilibrium by measuring its Planckian spectrum, with the 
temperature $T_0 = 2.725 \pm 0.002$ K~\cite{FIRAS-COBE}. 
Although the accuracy of these data (within limits as tight as 0.03\% in 
the frequency range 60--600 GHz) left no doubt about the past thermal 
equlibrium state of the CMB, it is still possible that such Planck law 
derived within Boltzmann-Gibbs statistics does not exactly describe this 
radiation and may instead obey a generalized expression. 
A plausible distribution could be that obtained within nonextensive 
statistical mechanics (for a review on this subject see, e.g.~\cite{Tsallis}), 
with values of the $q_a$-parameter ranging in the interval 
$|q_a - 1| \alt 5 \times 10^{-4}$ ($a$ stands for {\textit average} and 
refers to the quasi-Planckian distribution corresponding to $T_0$) as shown 
in~\cite{TBL}.

Observations with another instrument on board COBE, the Dif\/ferential 
Microwave Radiometer (COBE-DMR)~\cite{COBE}, detected for the first time 
that the CMB contains tiny variations around $T_{0}$, termed CMB temperature 
f\/luctuations $\Delta T$, at the level of one part in $10^5$ on large 
angular scales ($\sim 7^{\circ}$). 
Since the standard inf\/lationary cosmology predicts that these temperature 
f\/luctuations should be isotropic and Gaussian random, the COBE-DMR data 
motivated a number of analyses, although not so accurate due to the large 
angular resolution of the data, to test the statistical properties of 
the CMB (see, e.g.~\cite{Ferreira}).

Recently, highly precise and excellent angular resolution data from the 
Wilkinson Microwave Anisotropy Probe (WMAP)~\cite{WMAP} confirmed the 
existence of the CMB temperature f\/luctuations.  
The WMAP satellite observes the microwave sky in five frequency bands, 
K, Ka, Q, V, and W, centred on the frequencies 22.8, 33.0, 40.7, 
60.8, and 93.5 GHz, respectively. 
The corresponding CMB maps released by the WMAP team are pixelized in the 
HEALPix scheme~\cite{Gorski} with a resolution parameter $N_{side}=512$, 
which means that the celestial sphere is covered with $3,\/145,\/728$ 
equal-area pixels, with a pixel size of $\sim 7$ arcmin. 

These highly accurate CMB data have renewed concerns over the CMB 
statistical properties, and considerable analyses of the Gaussian hypothesis 
have been done~\cite{Komatsu}. 
Clearly, the study of such hypothesis must take into account the possibility 
that deviations from Gaussianity may have non-cosmological origins such as 
unsubtracted foreground contamination, instrumental noise, and/or 
systematic ef\/fects~\cite{Eriksen}. 
But they may also have cosmological origin, as being, for instance, 
the ef\/fect of cosmic strings on CMB~\cite{JS}. 
In a variety of analyses using dif\/ferent mathematical tools, and 
including foreground cleaning processes aimed to eliminate possible 
non-Gaussian contaminations, many evidences regarding deviations from 
Gaussianity in the WMAP CMB data have been recently reported~\cite{Copi} 
(see also~\cite{Copi05} and references therein).  

In what follows we shall perform the statistical analysis of the Q, V 
and W maps, after the application of the Kp0 mask to eliminate known 
foregrounds, in order to determine how much their distributions of 
temperature f\/luctuations deviate, if they do, from the Gaussian 
distribution. 
Then we discuss the possible account of such distributions according to 
the non-Gaussian temperature distribution emerging within nonextensive 
statistical mechanics, 
because gravitation is a long-range interaction and this kind of phenomenum 
seems to be conveniently studied in the framework of this theory~\cite{AT}. 
Definitely, the statistical significance of our results shall be supported 
by the analysis of substantial Monte Carlo CMB maps. 

Now, we brief\/ly introduce the basics of nonextensive statistical 
mechanics (see~\cite{aplic} for various applications). 
The probability distribution $P_q(x)$, as a function of the variable $x$, 
results from the optimization of the $q$-entropy defined by~\cite{Tsallis} 
\begin{equation} \label{eq.1}
S_q \equiv k\{1-\int dx \, [P_q(x)]^q \}/(q-1) \, ,
\end{equation}
with the constraints
\begin{eqnarray}
\int dx \,P_q(x)  &=& 1 \, , \label{norma} \\
\frac{\int dx\,x\, [P_q(x)]^q}{\int dx\,[P_q(x)]^q} &=& c_1 \, , \\
\frac{\int dx\,x^2\, [P_q(x)]^q}{\int dx\,[P_q(x)]^q} &=& c_2 \, .
\end{eqnarray}
From this, one straightforwardly obtains
\begin{equation} \label{eq.5}
P_q(x) = \frac{e_q^{-\lambda_{1q} (x-c_1) - \lambda_{2q} (x^2-c_2)} }
{\int dx \, e_q^{-\lambda_{1q} (x-c_1) - \lambda_{2q} (x^2-c_2)} } \, ,
\end{equation}
where $\lambda_{1q}$ and $\lambda_{2q}$ are related to the Lagrange 
multipliers, and where 
\begin{equation}
e_q^z \equiv [1 + (1-q)z]^{1/(1-q)} \, , \,\,\,\,\,\,
\text{for $[1 + (1-q)z] \geq 0$} \, ,
\end{equation}
while $e_q^z = 0$ otherwise.
Note that this distribution function $P_q$ is the 
solution of a non-linear Fokker-Planck equation~\cite{TB}.
One observes that eq.~(\ref{eq.5}) can be rewritten as 
follows 
\begin{equation} \label{eq.7}
P_q(x) = \frac{e_q^{-B_q (x-x_0)^2} }
{\int dx \, e_q^{-B_q (x-x_0)^2}}
= A_q \, e_q^{-B_q (x-x_0)^2} \, ,
\end{equation}
where the values of $B_q$ and $x_0$ are related to $\lambda_{1q}$ and 
$\lambda_{2q}$; $A_q$ is the normalization constant obtained in such a 
way that eq.~(\ref{norma}) is satisfied.

We shall now use this nonextensive distribution to 
account for the CMB temperature f\/luctuations distribution of WMAP data.
In this case we have $x=T$, $x_0=T_0$, and $B_q=B(\nu)$ (strictly 
speaking $B_q=B_q(\nu)$; however, for the cases considered here we 
have just one value for the $q$-parameter). 
Thus, 
\begin{equation}
P_q(\Delta T) = A_q \, e_q^{-B(\nu) \Delta T\:\!^2} \, , 
\end{equation}
where we wrote $P_q(\Delta T)$ instead of $P_q(T)$ to be clear that our 
analysis is dealing with the statistics of the temperature f\/luctuations. 
In the limit $q \to 1$, we recover the Gaussian distribution 
\begin{equation}
P_q \to P^{\text{Gauss}} = A\, e^{-B(\nu) \Delta T\:\!^2} \, , 
\end{equation}
where 
$A \equiv 1/(\sigma_{\nu}\sqrt{2\pi}), \, 
B(\nu) \equiv 1/(2\,\sigma_{\nu}^2)$, and $\sigma_{\nu}^2$ is the 
variance of the Gaussian distribution. 
For a comparison of the distributions $P_q$ with $P^{\text{Gauss}}$, 
in our analyses we assume $A_q = A$.

As recently pointed out by~\textcite{JS}, the signal measured at any pixel 
in the microwave sky is made of several components 
\begin{eqnarray}
T_{\mbox{\rm pixel}}=T_{\mbox{\rm foregrounds}}+T_{\mbox{\rm noise}}
                    +T_{_{\mbox{\sc cmb}}} \, ,
\end{eqnarray}
corresponding to foreground signals, the noise from the instruments, 
and the CMB temperature f\/luctuations, respectively. 
Foreground contributions, expected to be small away from the Galactic 
plane and with point sources punched out, are removed from the beginning 
with the application of the Kp0 mask~\footnote{The Q, V, and W maps here 
analysed were already corrected by the WMAP team for the Galactic 
foregrounds (synchrotron, free-free, and dust emission) 
using the 3-band, 5-parameter template fitting method described 
in~\cite{WMAP2}.
However, the foreground removal is only applicable to regions outside 
the Kp2 mask. 
For this, the Kp2 mask, or preferably the Kp0 mask, should be applied 
for the statistical analysis of the CMB maps.}.
Thus, one is left with CMB signal plus the Gaussian signal from the 
instrument noise (actually the signal noise is Gaussian {\it per} 
observation~\cite{JS,Jarosik}).

Thus, one has to consider these ef\/fects by defining~\cite{JS} 
the variance of the total signal $T_{\mbox{pixel}}$ as
\begin{eqnarray} \label{sigmaG}
\sigma_{\nu}^2 \equiv \sigma^2_{\mbox{\rm pixel}} = 
\frac{\sigma_0^2}{n_{i}} + \sigma_{\mbox{\sc cmb}}^2 \, ,
\end{eqnarray}
where $n_i$ is the number of observations for the $i$\/th pixel,
$\sigma_0^2$ is the noise variance {\it per} observation, and 
$\sigma_{\mbox{\sc cmb}}^2$ is the variance of the CMB temperature
f\/luctuations.
The mean contribution of the instrumental noise can be estimated by  
considering the ef\/fect of the dif\/ferent number of observations for each 
pixel (see Ref.~\cite{Jarosik}, pag. 16). 
Thus, given a CMB map, this can be done with the ef\/fective variance due 
to noise
\begin{eqnarray} \label{sigmaNoise}
\sigma^2_{\mbox{\rm noise}}(\nu) = \frac{\sigma_0^2 \sum^N_{i=1} (1/n_i)}{N}\, ,
\end{eqnarray}
where $\sigma_0^2$ is the variance {\it per} observation characteristic of 
the instrument (radiometer), 
$n_i$ is the ef\/fective number of observations for the $i$\/th pixel,
and $N$ is the total number of pixels considered in the analysis of the map. 
For the Q, V and W maps we obtained $\sum (1/n_i)= 7,\/423.90$, $5,\/465.98$, 
and $1,\/816.96$, respectively; moreover the number of pixels analysed $N$ 
is equal for the three maps $N_Q=N_V=N_W=2,\/414,\/705$. 
In other words, the ef\/fective noise variances $\sigma^2_{\mbox{\rm noise}}(\nu)$ 
and the variances $\sigma_{\nu}^2$ leads to the CMB variance 
\begin{eqnarray} \label{sigmaCMB}
\sigma^2_{\mbox{\sc cmb}} = \sigma_{\nu}^2 - \sigma^2_{\mbox{\rm noise}}(\nu)\,.
\end{eqnarray}
Although both $\sigma_{\nu}^2$ and $\sigma^2_{\mbox{\rm noise}}(\nu)$ depend 
on the map under analyses, their dif\/ference is independent of the map,
in other words, if our treatment of instrumental noise is correct, 
the CMB variance $\sigma^2_{\mbox{\sc cmb}}$ should be the same for 
the three CMB maps under investigation (Q, V, and W).

As previously reported (see section 2.2 and Fig. 2 in Ref.~\cite{JS})
the WMAP CMB temperature distribution does not fully obey a Gaussian 
temperature distribution 
\begin{eqnarray}
P^{\text{Gauss}}(\Delta T) = \frac{1}{\sigma_{\nu}\, \sqrt{2\pi}} \,
e^{-(1/2\sigma_{\nu}^2) \Delta T\:\!^2} \, .
\end{eqnarray}
In fact, the deviation from a Gaussian distribution can be apreciated in 
Figs. 1a, 1c, and 1e, for the Q, V, and W maps, respectively. 
In our analyses, the best-fit Gaussian to the data was obtained according 
to the $\chi^2$/degree of freedom (dof) estimator test. 
Thus, the $\chi^2 / 200$ values for the Gaussian fits are $0.116, 0.275$, 
and $0.167$, for the Q, V and W maps, respectively, and the $\chi^2 / 200$ 
values for the nonextensive temperature distribution $P_q$ are 
$0.00155, 0.00178$, and $0.00216$, for the Q, V and W maps, respectively. 
Analyses performed using 400 dof instead of 200 dof result in similar 
estimative values.  

In Fig. 1 the blue lines correspond to the best-fit Gaussian with 
variances $\sigma_{{\text{Q}}}=140.58 \mu$K, 
$\sigma_{{\text{V}}}=163.52 \mu$K, and $\sigma_{{\text{W}}}=190.35 \mu$K, 
while the red lines correspond to the best nonextensive distribution fit 
with the same variances and $q=1.056$.
Once the variances $\sigma_{\nu}^2$ have been determined through the 
$\chi^2$ best-fit Gaussian temperature distribution, for each of the CMB 
maps, then we use the ef\/fective noise variance $\sigma^2_{\mbox{\rm noise}}$ 
given in eq.~(\ref{sigmaNoise}) again for each of the CMB maps,
to calculate the CMB variance. 
Our results are 
$\sigma^2_{\mbox{\sc cmb}}=(68.77)^2, (69.34)^2, (68.81)^2 \mu\mbox{\rm K}^2$, 
for the Q, V, and W maps, respectively, in excellent agreement with what 
is expected. 
This result validates the ef\/fective noise variance as representing the 
mean contribution of the instrumental noise.

\noindent
In Figs. 1a, 1c, and 1e we plotted the temperature distributions 
of the Q, V, and W maps, in the form 
$\log_{10}(P^{\text{CMB data}}), \log_{10}(P^{\text{Gauss}})$, and 
$\log_{10}(P_q)$ {\textit versus} $\Delta T$ using the $\chi^2$ 
best-fit Gaussian temperature distribution $P^{\text{Gauss}}$ 
and also the best-fit nonextensive distributions $P_q$ with $q=1.056$.
To enhance the non-Gaussian behavior of the WMAP data, in Figs. 1b, 1d, 
and 1f, we plotted instead 
$\log_{10}(P^{\text{CMB data}}), \log_{10}(P^{\text{Gauss}}),$ and 
$\log_{10}(P_q)$ {\textit versus} $(\Delta T/\sigma_{\nu})^2$,
since linearity (blue curves) corresponds to the best-fit Gaussian 
distribution.
Our results evidence that the distibution of the CMB temperature 
f\/luctuations does not obey a Gaussian distribution. 

\begin{figure*}
\includegraphics[width=14.5cm, height=21cm]{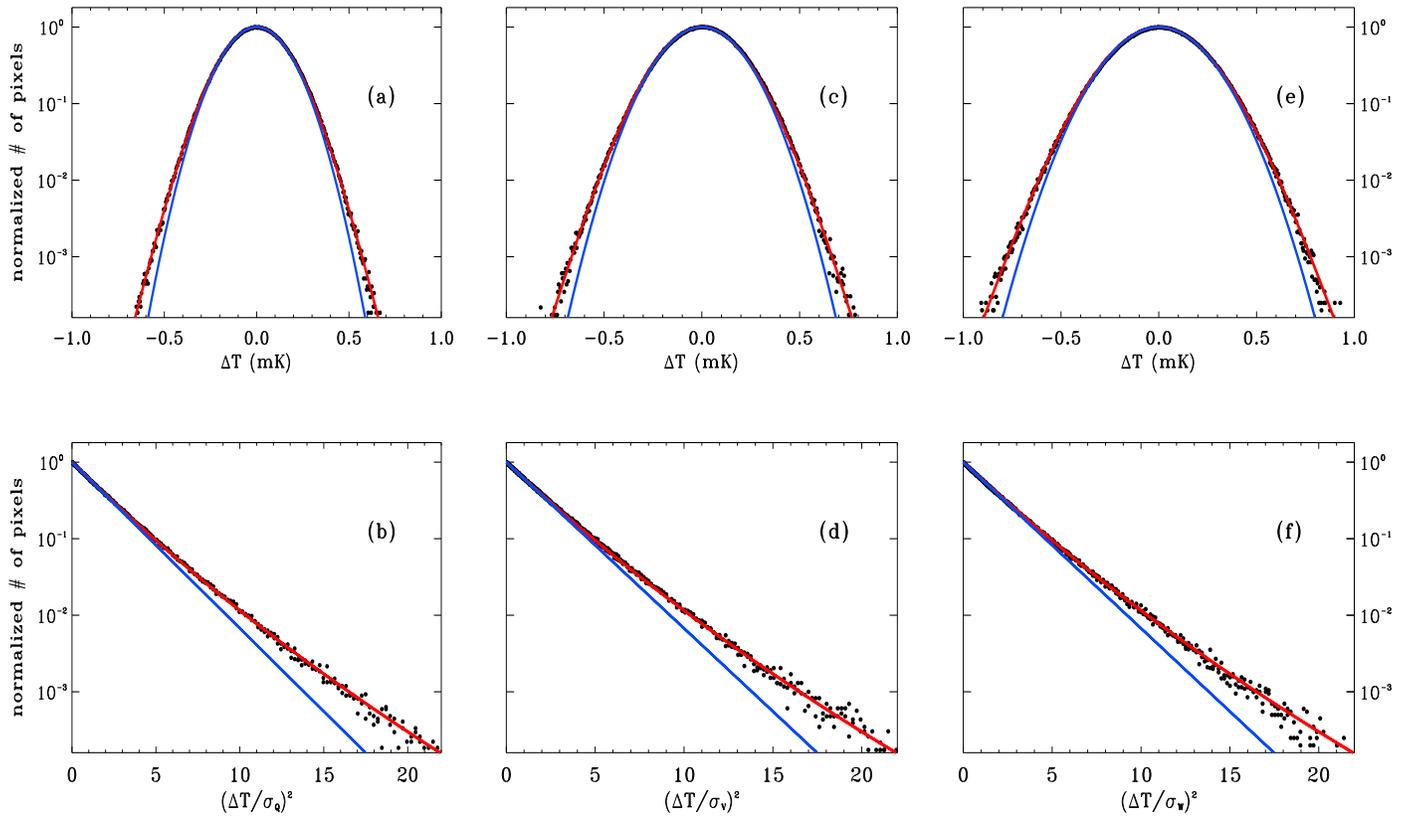}
\vspace{-9cm}
\caption{\label{figure1} Fits to CMB temperature f\/luctuations
measured by WMAP in bands Q, V, and W (after using the Kp0 mask)
applying an exact Gaussian distribution (blue curve) and a 
nonextensive function (red curve).
In figures (a), (c), and (e) we plotted the normalized number 
of pixels {\textit versus} $\Delta T$, while in figures (b), (d), 
and (f) we plotted the normalized number of pixels {\textit versus} 
$(\Delta T/\sigma_{\nu})^2$, respectively. 
The $\chi^2$ best-fit for the distributions $P_q$ gives $q=1.056$, 
with $\sigma_{{\text{Q}}}=140.58 \mu$K, $\sigma_{{\text{V}}}=163.52 \mu$K, 
and  $\sigma_{{\text{W}}}=190.35 \mu$K, respectively.}
\end{figure*}

\begin{figure*}
\includegraphics[width=14.5cm, height=21cm]{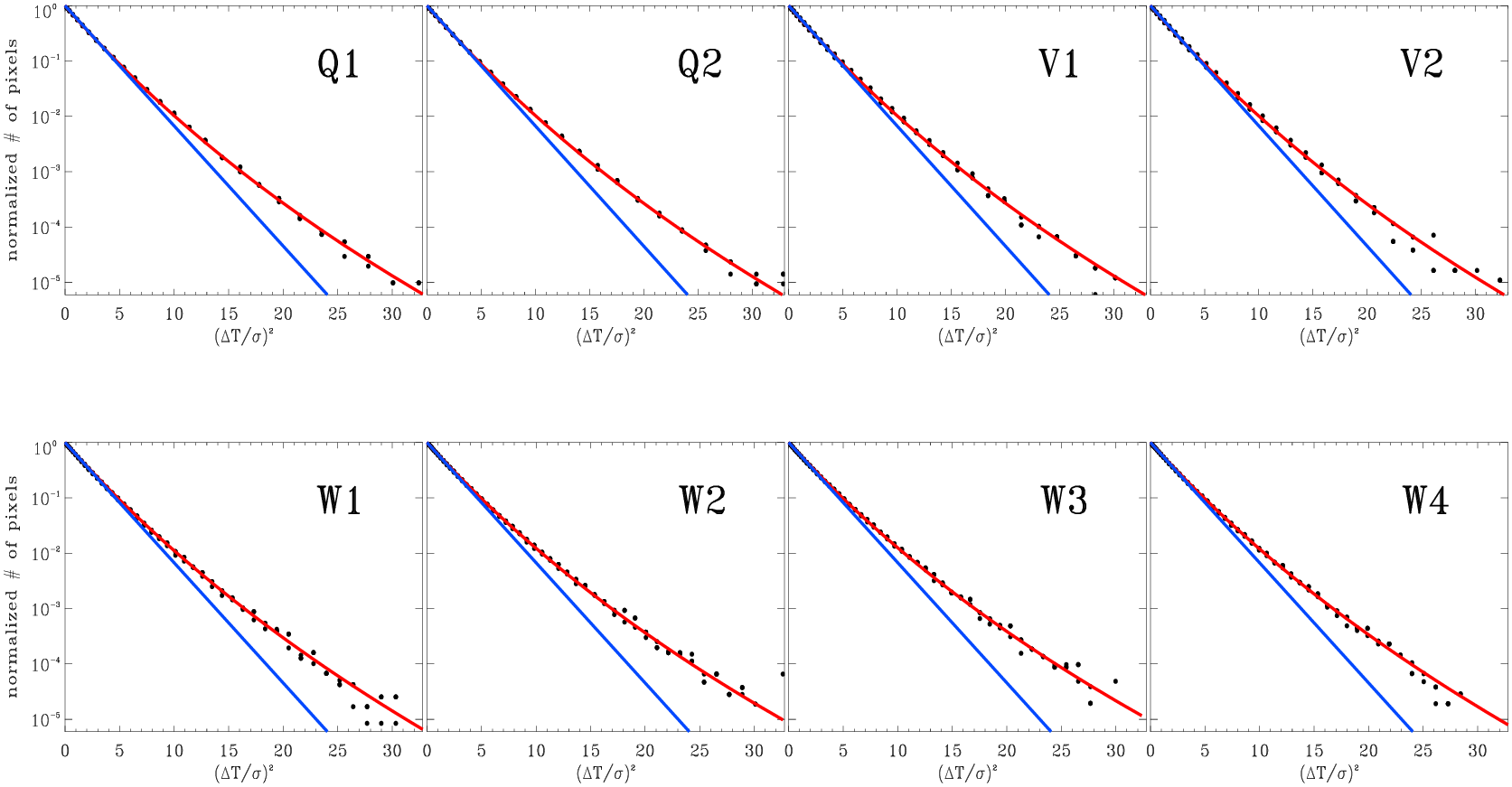}
\vspace{-9.5cm}
\caption{\label{figure2} Results of the analysis of WMAP data using 
the individual radiometer maps Q1, Q2, V1, V2, W1, W2, W3, and W4, 
after applying the Kp0 mask. 
In all plots the blue curve represents the best-fit Gaussian, according 
to the $\chi^2$ estimator, while the red curve represents the best 
nonextensive distribution function. 
Each $P_q$ distributions fits the corresponding data with a slightly 
dif\/ferent value of $q$, where the mean value of them is 
$q=1.055 \pm 0.002$.}
\end{figure*}

In order to strengthen our analysis, we removed fits to the WMAP foreground 
templates for the eight individual radiometer maps (corresponding to 
radiometers Q1, Q2, V1, V2, W1, W2, W3, and W4), after applying the Kp0 
mask to avoid high latitude foregrounds. 
Our results are plotted in Fig. 2. 
As observed, these results were essentially the same as those shown in 
Fig. 1, that is, $q=1.055 \pm 0.002$. 

Then, we investigated the possibility that the discrepancies observed 
between WMAP data and exact Gaussian distributions, as evidenced in Figs. 
1 and 2, occur just by chance. 
For this scope, we analysed a set of $10,\/000$ Monte Carlo realizations 
of CMB Gaussian maps, and we found, at the 99\% confidence level (CL), 
that the distributions of CMB temperature f\/luctuations measured by WMAP 
are not properly described by exact Gaussian temperature distributions.  

We also investigated, through substantial numerical simulations, the 
possibility that the instrumental noise could introduce non-Gaussian 
signals in the CMB maps.
According to~\cite{Jarosik,JS} instrument noise produces a random Gaussian 
signal, actually Gaussian {\it per} observation. 
Since pixels are observed a dif\/ferent number of times, the ef\/fect on 
the CMB map could be a significant non-Gaussian signal in the CMB temperature 
distribution, and clearly this possibility should be taken into account in 
the simulations.
For this, to each one of the $10,\/000$ Monte Carlo CMB Gaussian maps we 
added a simulated non-stationary Gaussian radiometer noise, taking into 
account the actual number of observations ($n_i$) for each pixel in the 
maps. 
We consider the $n_i$ data from the WMAP-W4 map~\cite{WMAP}.
We obtained $q=1.005 \pm 0.01$ for the CMB plus non-stationary Gaussian 
noise simulated maps, which is obviously consistent with Gaussianity. 
A $\chi^2$ estimator test shows that the $P_q(\Delta T)$ does not fit 
the simulated data as well as it does the WMAP data. 
The monotonic discrepancy behavior observed in Figs. 1b, 1d, and 1f, 
which is well fitted by the nonextensive expression given in eq.~(\ref{eq.7})
due to a minimum $\chi^2$ value, was observed in less than 1\% of the 
simulations performed. 
Therefore we can rule out, at the 99\% CL, the simulated radiometer noise 
as being the explanation for the non-Gaussianity observed in the WMAP 
temperature f\/luctuations.

In conclusion, we have shown that an exact Gaussian distribution is 
excluded, at the 99\% CL, to properly represent the CMB temperature 
f\/luctuations measured by WMAP. 
Although the value of the $q$-parameter is close to 1, our analyses indicate 
that to consider these temperature f\/luctuations as being of Gaussian nature 
is not rigourously exact and it should be considered as a good approximation 
instead.

We acknowledge use of the Legacy Archive for Microwave Background Data 
Analysis (LAMBDA). 
C.T. acknowledges the partial support given by Pronex/MCT, CNPq and FAPERJ 
(Brazilian Agencies). 
T.V. acknowledges CNPq grant 302266/88-7-FA and FAPESP grant 00/06770-2.
A.B. acknowledges a PCI/DTI/7B-MCT fellowship. 
We thank Carlos A. Wuensche for his unvaluable help in the Monte Carlo 
analyses. 
Some of the results in this paper have been derived using the 
HEALPix~\cite{Gorski} package.

\end{document}